# India Attempts to Give a Jump-start to Its Derailed Telecommunications Liberalization Process

**Introduction**

This paper takes India's case as an example to show that a need for a proper legal and regulatory regime at an institutional level at the outset, and a clear commitment to pro-competitive market principles at the political level, are necessary pre-conditions to successfully reforming the telecom sector.

After the 1991 Economic Policy made a shift from a closed economic model to a market-oriented model, private sector was invited to participate in reforming India's telecom sector. However, the government took a half-hearted approach in overhauling the legal and regulatory regime. Competition was allowed in cellular and basic services. The Ministry of Communications and the incumbent Department of Telecommunications (DOT) issued licenses to their competitors. Lack of transparency in issuing licenses and unrealistic license fees derailed the reform process and led to wasteful litigation. The courts did not support the regulator and virtually made its role redundant.

The paper discusses the various domestic and international factors that influenced the government to take initiatives to overhaul the legal and regulatory regime to untangle the problems of its *1994 Telecom Policy* and make it more compatible with the new convergence era.

*The Internet Policy of 1998* will be discussed to show how competition in Internet services and infrastructure is impacting the telecom sector in fulfilling public policy goals. The objectives of the *1999 Telecom Policy* will be discussed to show that the policy took a positive direction. It simplified the licensing issues and also allowed more competition in cellular, basic and long-distance services. However, the policy did not clarify the regulator's role, and the government retained tools to control the future of the liberalization process.

After the elections in 1999, the telecom sector was the key sector in which major decisions were made. The DOT's service arm was corporatized and the government decided to privatize its international long-distance carrier, Videsh Sanchar Nigam Ltd. (VSNL), before 2002. It decided to introduce more competition in cellular, basic and domestic long-distance services.

The regulator's role has been clarified and its regulatory function has been separated from its dispute settlement and a new dispute settlement body has been created. The government is debating whether to replace the existing legal statutes with a common statute to regulate broadcasting, telecommunications and information technology sectors. The statue would create a common regulator and give it overall responsibility for a range of matters, from issuing licenses to settling disputes. All the steps taken by the government sent positive signals in reviving the industry participants' confidence. It also shows that India has come a long way from a closed, centralized regime to a more decentralized regime. The government also seems to be more committed in this round.

**The Indian Telecommunications Sector**

Until 1985 India's telecom sector was a typical PTT model. The Ministry of Posts and Telegraphs and its department controlled the posts and telecom services and infrastructure. The sector was mainly governed by the *Indian Telegraph Act of 1885*. In 1985 postal services were separated from telecom, and the Ministry of Communications and its Department of Telecommunications (DOT) were assigned powers including manufacturing equipment, policymaking, providing services and creating infrastructure.

However, telecommunications in India was *not* considered a necessity like in North America; rather, it was luxury to have a telephone connection. The government did not give priority to the development of this sector. The national five-year Plans show that until 1985, the government's investment in the telecom sector was less than three per cent; it was only from 1992 onwards that government spending increased to 11.9 per cent and more.[1] Therefore, DOT, lacking funds, focussed on spreading services and infrastructure mainly in major cities. The quality of service was poor and there were huge waiting lists to get a telephone connection. The government used this sector to fulfill its political goals of providing employment. As a result, DOT was bloated with huge labour and civil servant unions (470, 000 employees).

**Initiatives to Upgrade the Industry**

It was only in the early eighties, with technological advancements and international developments, that Prime Minister Rajiv Gandhi took initiatives to move from the closed-economy ideology towards more market-oriented reforms. He brought a fresh promise of market reforms in the information technology (IT) and telecommunications industry. The information technology policy focussed on the export of software and computers. To achieve this goal, tariffs were lowered on imported goods and direct access to imported equipment and technologies was allowed. To facilitate software exports, "Software Technology Parks" were created in selected cities and exporters were allowed to set up satellite communications. In these parks other infrastructure facilities such as buildings, electricity and high-speed data links were provided. Thus, partnership between the Department of Information Technology and the private sector worked well for achieving public policy goals. It is reported that the IT industry in 2000-01 has crossed $ 10 billion mark; its earning are mainly from software services, exports, and the computer training industry.[2] A recent article in the Wall Street Journal confirms that India's growing success and interest in the information technology industry, both at home and abroad, has led the government to partner with the Massachusetts Institute of Technology (MIT) to create an Asian Media Lab to take new initiatives in addressing very fundamental problems of literacy, health, and entrepreneurship. [3]

However, in the case of telecommunications, the government's approach was half-hearted. The government was reluctant to break the dominance of DOT or let go of its own powers. In 1984, overcoming resistance from DOT, the government was successful in allowing private investment in the manufacture of customer premise equipment. The government franchised public call offices and also set up two new companies. Mahanagar Telephone Nigam Limited (MTNL), a mostly government owned company which offered basic services in Delhi and Bombay, and Videsh Sanchar Nigam Limited (VSNL), which offered international long-distance services. [4] Although these initiatives proved to be positive, they could not bridge the overwhelming demand for infrastructure and services.

Advancement of technologies, international actors such as the Wold Bank and trade blocks (OECD), domestic industry and consumers continued to point out to the government that the telephone was a necessity and that lack of infrastructure was the root cause for the economic slowdown.[5] The government responded by constituting independent telecom reconstruction committees (for instance, Artherya committee) for taking advice.[6] Generally, the committee's recommendations were pro-competitive; however, the politically influential DOT was successful each time in diluting the recommendations in its favour.

---

[1] N. Ravi "Telecommunications in India-Past, Present and Future" (March 1992) *IEEE Communications Magazine*.

[2] "Indian IT Industry Crosses $10-B Mark" The Economic Times (12 July 2001) available at http://www.economictimes.com/120701/12tech17.htm

[3] David Armstrong & Jesse Pesta "MIT, India are Close to Asian Media Lab Pact" *Wall street Journal* (12 February 2001)

In 1992, competition was allowed in cellular mobile services. However, the government failed to overhaul the legal and regulatory regime to suit the competition process, and the antique *1885 Act* was left in place to govern the new competitive era. The Ministry of Communications and the DOT became responsible for issuing and regulating licenses to new entrants in Delhi, Bombay (Mumbai), Madras (Chennai) and Calcutta. Two private operators were allowed to compete in these four major cities. Before the licenses were issued, there was flurry of petitions before the courts questioning the procedure adopted by the selection committee to evaluate the tenders. The courts showed judicial deference with respect to technical and economic issues and adopted an essentially "hands off" approach.[7]

Although the liberalization process had a rough start, it was clear that the Indian market was huge and therefore had great potential to draw private investment. The DOT, in May 1993, asked the leading Indian financial institution, Industrial Credit and

---

[4] The government holds 56 per cent equity of MTNL and 52 per cent of VSNL.

[5] For more information on the debates among various domestic and international actors see, D.M. Stephan, *Globalization, Liberalization and Policy Change, a Political Economy of India's Communications Sector* (New York: St. Martin's Press, 1997).

[6] Arthreya M.B "India's Telecommunications Policy, A Paradigm Shift" (1996) Vol.20 No.1 *Telecommunications Policy* at 11.

Investment Corporation of India (ICICI) to prepare a report on the development of regulatory institution and to recommend terms and conditions for the private sector entry in basic services.

The ICICI noted:

> In an emerging economy, great amounts of discretion lie with the sector managers of the government, which, in India, lies with the Ministry of Communications [in practice, coextensive with the DOT]. If such discretion is unlimited or unrestrained and vests in a single person or office at any time, it is unlikely to command the level of comfort which is mandatory for attracting substantial private investment. The discretion should be so hedged that the decision-making process becomes tolerably predictable and reliable…and transparent. The investor would rather go by the written law… than be guided by platitudes and pronouncements of the decision-maker. Thus, until the law is formally changed…the value of the investment opportunity will be dismissed. [8]

In any event, this argument was to be disapproved. Private investors, both domestic and foreign, showed great enthusiasm to invest in India. As Mody pointed out, "the government instead of confronting the fundamental differences of interest and ideology within the telecom policy compromised by announcing the *National Telecom Policy of 1994*.[9]

**The Telecom Policy 1994**

The *1994 policy* totally dismissed privatization of incumbent companies and made it clear that the much- needed private foreign investment will supplement DOT's efforts in spreading basic telephony.[10] The policy was unrealistic as it failed to clearly define how universal access and service goals were to be achieved. On one hand, it admitted that India had 0.8 per cent of teledensity; on the other, it stated that the telephone would be provided on demand by 1997. It allowed competition in basic services by way of duopoly in each state (called circles) and allowed foreign companies to partner with Indian counterparts to compete with DOT in 20 circles. However, implementation of the policy was left in the hands of the incumbent DOT. DOT's bureaucrats, fearing competition, used their licensing and regulatory powers and distorted the policy in their favour.[11]

---

[7] *Tata Cellular* v. *Union of India* (1996) A.I.R 12-50 (Supreme Court of India)

[8] See *ICICI's Report on the Telecom Regulatory Body for India* (New Delhi, January, 1994), *Report on Entry Conditions for Basic Telecom Services* (New Delhi, June 1994).

[9] Mody. B "State Consolidation Through Liberalization of Telecommunications in India" (1995) Vol. 45 *Journal of Communications* at 107-123.

In August 1995, the first eight cellular licensees in four metropolitan cities began commercial services. In late 1995, the DOT and the Ministry issued about three dozen licenses to cellular operators in all other provinces of India. August 31,1995, was supposed to be the day to end DOT's monopoly over basic services. But it became a day of shock and surprises when new entrants legally challenged competition in basic services. In *Delhi Science Forum and Others v. Union of India and Another* tenders were requested from private companies to compete with DOT/MTNL for provision of basic services in different states divided into 20 circles.[12] During the evaluation of tenders, a new company, Himachal Futuristic Communication (HFCL) offered the highest bid for nine circles. The Ministry of Communications was uncomfortable with the idea of handing over nine circles to one company. Thus, the ministry devised new rules for capping the system and gave HFCL the choice of any three circles among the nine circles on which it bid. As a result, the validity of the procedures adopted by the government and DOT was challenged by competitors. DOT had armed itself with a clause carefully drafted in its favour by which it was free to restrict the number of service areas for which one company can be licensed to provide service. On the basis of this clause, DOT escaped legal scrutiny.[13] This incident, however, became very controversial. This time, the courts recognized that there was a need for an independent regulatory body to deal with complicated economic and technical issues. Nonetheless, they felt incapable of forcing the government to replace the *1885 Act* with legislation more suitable for competition.

---

[10] See Government of India, Press Information Bureau, National Telecommunications Policy 1994. The NTPs is available at http://www.trai.gov.in/policy.htm

[11] Interconnection charges were in-sided, high license fee to be paid at the beginning of the year, entrants were not allowed to connect adjacent circles to provide long-distance services, they could not directly connect with VSNL, restrictions were on usage of technology and licenses were not allowed to be transferred. See Chowdary T.H "Liberalization of Indian Telecommunications: Causes, Course and Critical Issues" (1997) Global Networking '97 (Calgary, 1997) at 101.

[12] See *Delhi Science Forum and Others* v. *Union of India and Another* (1996) 2 A.I.R. 1356-1368 (Supreme Court of India)

The controversy and delay in awarding basic service licenses led only six companies to take licenses in offering services in only few states.[14] For the remaining states there was no company ready to match the unrealistic reserve- price, which the ministry and DOT had fixed. Moreover, the companies who had taken licenses later failed to fulfill their license conditions as their business plans were flawed. They also had serious claims against the government and its agencies, especially DOT, for not giving timely clearances in order for them to start their commercial operations on time. Both cellular and basic service operators had committed to unrealistic license fees and were struggling

---

[13] For details of the bidding patterns in basic service competition see Dokeniya. K "Re-forming the State: Telecom Liberalization in India" (1999) Vol.23 No.*2 Telecommunications Policy* at 112-115.

[14] In 1997, the government issued first license to Bharati Telenet to offer services in Madhya Pradesh and then to Tata Teleservices for Andhra Pradesh. Reliance Telecom was given Gujarat and Essar for the state

to survive in the Indian market. They owed almost $873 million to the government towards their outstanding license fees. By 1999, only two basic service operators were able to offer services in two provinces and they had few subscribers; the cellular operators had signed only one million subscribers. Therefore, to a certain extent, these companies, in going along with government's ineptitude to some extent were responsible for delaying needed reforms.

While the controversy over basic service competition was going on in India, the government in the WTO's Basic Telecommunications Service Agreement and its Reference Paper took a very protectionist stance. For instance, in its domestic policies it allowed foreign investment up to 49 per cent; however, in the WTO, it decided to retain 25 per cent.[15] The basic service controversy ultimately led the government to promulgate the *Telecom Regulatory Authority of India Act* in 1997 (TRAI Act). This Act established the independent regulatory body Telecom Regulatory Authority of India (TRAI).

**The Telecom Regulatory Authority of India Act, 1997 (TRAI Act)**

The enactment of the TRAI Act was also a long delayed legislative process. In the view of an independent consultant who participated in the drafting exercise, Mahesh Uppal:

> …to suggest, even indirectly, that the government had, at that time [of drafting the TRAI Act], a clear idea of what it was that it wanted the TRAI to achieve, is stretching credibility. India's telecom regime was schizophrenic during the period of drafting and redrafting of the TRAI Act and indeed much of the period before that. At virtually every stage of the process, there were controversies. There were frequent spats in high levels

---

of Punjab. Hughes Ispat Ltd for Maharastra. In 1998 Shyam Telelink Ltd was given license for the state of Rajasthan. The list is available at http://www.abto.org

[15] For India's commitments at the WTO see http://www.wto.com

and sudden changes of guard at Sanchar Bhavan [DOT headquarters]. There was no evidence of focussed policy making or implementation.[16]

The creation of the new independent regulatory agency was a significant episode, perhaps ultimately a defining one. TRAI was not given responsibility to issue and revoke licenses, but only to recommend them. It had responsibility to fix tariffs and resolve disputes. The DOT surrendered its regulatory role in principle, though it still retained policy-making, licensing, and operative powers within the same organizational boundaries. But it soon became evident that DOT was by no means prepared to defer the regulatory powers to TRAI.

**Legal Battle between TRAI and DOT**

From the moment TRAI was established it was thrust into an uncomfortable role. It had to take control of a fledging private industry experiencing a sluggish reception from the market and an unsympathetic incumbent who used all the resources to defend its territory. In principle, TRAI was to regulate the industry, but in practice it would have to attempt, despite itself, to affect some reform of DOT that the political class had never had the courage to undertake.

The newly appointed TRAI was plunged into a bitter confrontation between DOT and the cellular operators. TRAI gave relief to the cellular operators who had alleged that DOT had unilaterally increased the tariffs for calls made from ordinary telephone to cellular-mobile phone.[17] DOT took this reversal badly. It appealed the TRAI order to the courts, inviting the Delhi High Court to reconsider the case in its entirety as well as to trim the TRAI's scope and powers. This was the first of a series of bruising collisions between the regulator and the incumbent. On one count, within a year between a dozen and twenty of these cases were awaiting judicial clarification of the jurisdictions of TRAI and the DOT.

---

[16] Personal interview with Mr. Mahesh Uppal in Delhi on October 23 1998.

[17] *Aircel Digilink India Limited and Others* v. *Union of India and Others* (25 April 1997) (The Telecom Regulatory Authority of India, New Delhi)

Another highly contentious petition was that between *M/s Bharti Cellular Ltd & Another v. Union of India*. Entrants challenged DOT and the government with respect to a unilateral decision to introduce MNTL to offer cellular services and framing an Internet Policy without the recommendation of TRAI. They also challenged DOT's threatened revocation of licenses without the recommendation of TRAI.[18] In all the three complaints the crucial point to be considered was whether it was mandatory for DOT and government to seek TRAI's recommendation before introducing a new provider, setting terms and conditions of license to a service provider, or revoking a license of a service provider for non-compliance thereof, or whether the *TRAI Act* was merely a means to enable TRAI to express its opinion with regard to these matters.

While deciding these complaints, TRAI explained at length the relationship between a comprehensive telecom environment and an independent regulation and ruled that it was mandatory for the government and DOT to have before them its recommendation with regard to matters covered under the said provisions. The DOT was predictably unhappy at this decision and concluded that TRAI was systematically biased against the publicly- owned operators.

---

[18] *M/s Bharati Cellular Ltd & Another v. Union of India & Another* (17 February 1998) (The Telecom Regulatory Authority of India, New Delhi)

The relations between the two agencies had become so notoriously bad that the Prime Minister's office intervened to reconcile the two.[19] However, senior officials were not ready to risk the impact on investors of amending the TRAI's constitutive legislation to suit the DOT, nor were they, despite public appeals by the Chairman of TRAI, Justice S.S. Sodhi, willing to cut through the litigation between two agencies of the central government by restructuring the DOT or strengthening the *TRAI Act of 1997*.

The lack of response from the political executive to either the DOT's proposal to cut the TRAI down to size, or to Sodhi's appeal to confirm the TRAI's contested mandate, left the court as the only avenue, which promised an eventual resolution. After the TRAI order blocking MTNL from proceeding with its cellular plans, the DOT appealed the TRAI's decision to the Delhi High Court and asked for an interpretation of the functions of the TRAI and their implications for the DOT. The High Court's decision heartened the DOT.

Justice Usha Mehra found it difficult to reconcile the *TRAI Act 1997*, and its expression of the powers of a quasi-judicial regulatory agency, with her understanding of a role of the executive: "it is unimaginable that the power to grant license rests with the government but would be subject to the discretion of another Authority." This situation would make the DOT's powers under the *Indian Telegraph Act of 1885* "redundant" and "this court is not in a position to declare [that]." She set aside TRAI's decision.[20] This judgement was criticized on legal grounds for its failure to try to reconcile the two pieces of legislation. On appeal, the Division Bench of Delhi High

---

[19] Navika Kumar "PMO Initiates TRAI, DOT Patch-up" *The Indian Express* (30 April 1998)

Court upheld the decision of Justice Usha Mehra and virtually made TRAI's role "redundant".[21]

In spite of not having support from the government, the incumbent, or the courts, TRAI was pro-competitive and was successful to a certain extent in rebalancing telecom rates.[22] TRAI also went ahead with the consultative process with various aspects of the telecom industry to introduce more competition.

However, disillusioned with the Indian government's bad handling of telecom deregulation, several multinational companies started to pull back their investments and blamed the government for having "unfriendly telecom policies."

The Telecom sector provided $5 billion in foreign direct investment, one of the highest in the Indian economy, since the market opened in the early 1990s. As a result, the government had little choice but to take steps to untangle the problems.

**Attempts to Jump-start the Derailed Liberalization Process**

**Internet Policy -1998**

As discussed above, the government was successful in liberalizing the information technology sector. Recognizing the explosive growth of the Internet, it decided to formulate a new Internet policy by inviting various stakeholders to participate in the consultative process. The policy proposals were also posted on the Internet for recommendations. In 1998, the Internet policy was announced. It ended the monopoly of VSNL over Internet services and allowed unlimited competition with practically no license fee. The Policy also included certain telecommunications issues. Internet service providers (ISPs) could link their users by telephone lines or cable lines or even build their own transmission networks. Thus, the cable industry and the ISP's started to directly compete with the telephone companies in offering multi-media services. DOT issued more than 400 licenses including national, state and city licenses. More than 100 foreign investors have invested in these companies. With huge demand for bandwidth, domestic firms in partnership with international companies are actively taking part in competing with VSNL, MTNL and DOT to provide infrastructure and services. With legal and regulatory flexibility, private ISP's are now allowed to create, own and share private satellite based gateways and submarine cable landing stations to connect with international Internet gateways.[23] Indeed, competition has already made a substantial difference in the pricing of Internet services and infrastructure. According to a recent article in the Economic Times[24] by the year 2000, India had four million Internet users and by 2005, users are expected to reach 40 million. The report states that by 2005 India would have the second-largest Internet users' market in the Asia-Pacific region (following China). About 3.8 per cent of country's population would use the Internet.

---

[20] *Union of India v. Telecom Regulatory Authority of India and Others* (23 July 1998) Delhi High Court, Usha Mehra, J.

[21] *Microwave Communications Limited v. Union of India and Another* (12 October 1999) Delhi High Court, Arun Kumar J, & Manmohan Sarin j.

[22]The important areas of regulation were prices of local, long-distance and international calls. It was always alleged that DOT priced local calls below cost and subsidized local calls from its long-distance market. The TRAI embarked on a tariff rebalancing exercise meant to align prices to cost. It significantly reduced the prices of long-distance and international calls while increasing the cap on rentals and local calls. Thus the ability of the DOT to cross subsidize was eroded to some extent. The DOT after much infighting

**New Telecom Policy of 99[25]**

---

implemented the tariffs for long-distance and international calls but did not increase the rentals and prices of local calls to the extent allowed by the price caps.

[23] For the list of companies who have were issued licenses and those who have started operating to provide services and infrastructure see. www.dotindia.com

[24] See "Indian Net users-base to Touch 79-lakh by end-2201" *The Economic Times* (1 August 2001) available at http://www.economictimes.indiatimes.com/today/in01.htm

On March 27, 1999, the New Telecom Policy was announced. There was little hope that the new policy would bring profound changes as DOT members dominated the committee advising the government on telecommunications. Unexpectedly, the policy took a positive direction and was more detailed then its predecessor. It realistically targeted to achieve teledensity of 7 by 2005 and 15 by 2010. To increase rural telephony it targeted from 0.4 to 4 by year 2010. It also allowed more competition in basic, cellular and long-distance services. It confirmed that ISPs, cable companies, and public sector companies would be allowed to compete in various services. Furthermore, new entrants were allowed to pay a one- time entry fee and a license fee based on a revenue sharing scheme. To achieve universal service goals, resources were to be raised through a "universal access levy" from revenue earned by the operators. This revenue would be used in achieving universal service goals on TRAI's recommendations.

Although the policy did not resolve the problem of existing operators who owed million of dollars to the government towards licensing fee, on recommendations from the Attorney General of India, the government shifted all operators to a fixed entrance fee and rest by a revenue sharing scheme. Through this "bail-out package," the license period was also extended. In return, operators had to give up their duopoly status and withdraw litigation against the government.

The negative aspects of the policy included its failure to take a stand on the privatization of incumbent companies; in addition, it only promised to corporatize DOT by separating its service arm from its policy and licensing role. On one hand, it committed to a level playing field for all operators; on the other, it recognized the importance of the incumbent DOT and promised to reimburse full license fees on the grounds that DOT's had an obligation of providing telephony in rural areas. The policy failed to clearly define TRAI's role. It only briefly noted that the *1885 Act* need to be replaced. The government, in its sovereign capacity, retained the authority to intervene in the future competition process.

---

[25]New Telecom Policy of 1999 is available at dotindia.com

**Recent Developments**

Keeping the forward momentum, after the 1999 national elections, telecommunications was the key sector in which major decisions were made. The government on December 13, 1999 created a committee called the "Group on Telecom and Information Technology Convergence" (GTC) under the Chairmanship of Finance Minister Yashwant Sinha.[26] Mr. Sinha constituted three sub- groups to deal with various issues relating to telecommunications and IT. The sub-groups had to (a) consider and make recommendations to strengthen the TRAI through suitable legislative amendments, (b) identify and recommend measures for the resolution of subsisting problems for expeditious implementation of NTP-99, (c) to prepare the draft of a comprehensive statute to replace the *Indian Telegraph Act, 1885* keeping in view the rapid convergence of telecom, IT and broadcasting.

**Overhauling Legal and Regulatory Framework**

**Amendments to the TRAI Act of 1997**

Soon after the GTC was formed and the sub-groups were created, they sprang into action to overhaul the legal and regulatory regime and to expedite the implementation of the 1999 policy. On the recommendations of the sub-group headed by Information and Broadcasting Minister Arun Jailey to amend the TRAI Act of 1997, the parliament amended the Act in March 2000. TRAI's regulatory role was split from dispute settlement. The new TRAI, with two full time members and two part time, is responsible for recommending the introduction of new service providers, technological improvements, quality standards, and fixing the terms and conditions of licenses. It is mandatory for the government to seek TRAI's recommendations, although it will not necessarily accept them. The newly constituted Telecom Dispute Settlement and Appellate Tribunal (TDSAT) is empowered by a chairperson to have a one or two member bench to adjudicate disputes between a licensor and a licensee, service providers and between a service provider and a group of consumers. Decisions of the Tribunal can be challenged only in the Supreme Court. [27]

---

[26] The other members of the GTC are Ram Vilas Paswasn, Minister of Communications, Pramod Mahajan, Minister of Information Technology, Sushma Swaraj, Minister of Information and Broadcasting, Arun Jaitley, Minister of Law, Justice &Company Affairs, Fali Nariman, Member of Parliament, N. Vittal,

**The Communications Convergence Bill 2000**[28]

The sub-group headed by Fali Nariman, a Member of Parliament, put forward a detailed proposal of the Communications Convergence Bill 2000 and invited all the interested parties to make recommendations in replacing the *Indian Telegraph Act of 1885* and various other legislation relating to telecommunications, broadcasting and information technology. The Bill is posted on the Internet for submissions. Members of the committee are hoping to introduce the Bill in Parliament in the upcoming session for its approval.

---

Central Vigilance Commissioner, Dr. Montek Singh Ahluwalia, Member of Planning Commission, N.K. Singh, OSD, Prime Minister's office, Shyamal Ghosh, Secretary, Department of Telecommunications.

[27] Details of the amended Act are available at http://www.trai.gov.in

[28] The Communication Convergence Bill, 2000 is available at http://wwwidot.com

Spelling out the objectives, the Bill suggests that the Act may be called The Communication Convergence Act, 2000, which would establish an independent commission known as the Communications Commission of India (CCI). The Commission shall consist of a chairperson and seven members appointed by the government for five years from various specialized fields such as broadcasting, telecommunications, information technology, finance, management and law.[29]

The objectives of the Bill are to develop the communication sector in a competitive manner, and to ensure that market dominance in a converged environment is suitably regulated. In addition, the Bill proposes that communication services be made available at affordable cost and suggested that the Commission promote quality, plurality, diversity and choice of services. The Bill also aims to establish a modern and effective communications infrastructure, which takes into account the convergence of information technology, media, telecom and consumer electronics. As well, the Commission should promote equitable, non-discriminatory interconnection across various networks and promote an open licensing policy and, subject to spectrum, allow unlimited competition. It should also promote a level playing field among all operators.

The functions of the Commission are to facilitate and regulate all matters relating to the carriage and content of communications. The Commission will grant licenses and enforce license conditions and determine fees. It will determine appropriate tariffs and rates for licensed services. It will also have to ensure that the grant of licenses does not eliminate competition or lead to dominance by one or two players. It will formulate and determine conditions for fair, equitable and non-discriminatory access to network infrastructure facilities, protect consumer interests, and promote and enforce universal service obligations.

---

[29] On the recommendations of the Minister of Information and Broadcasting Minister Sushma Swaraj, the Group accepted that under CCI, for content regulation, members should include from art, literature,

The Commission will decide disputes between providers and a group of consumers in enforcing the provisions of the Act. It will decide disputes relating to spectrum interference, interconnectivity, denial of fair access and practices restrictive of fair competition. The Commission for any aggrieved person to appeal its order or decision will establish communications Appellate Tribunal. Only in exceptional cases can an order of the Appellate Tribunal be appealed to the Supreme Court of India.

The Commission will assign spectrum for commercial purposes to various users; however, the central government shall be responsible for the allocation of spectrum to the Commission, and the government can notify, the class of persons or services to which the spectrum is to be assigned. The government can also issue policy directives keeping the objectives of the Act in mind, and its decisions will be final.

**Competition in Services and Privatization of Incumbent Companies**

While the legal and regulatory framework is being overhauled, the group responsible for implementing the objectives of the *1999 Policy* made several decisions after taking recommendations from TRAI in allowing competition in various services and privatization of incumbent companies.

**Decisions for Reforming State Owned Companies**

In 1999, the government separated the DOT's service role from policy and licensing by creating the Department of Telecom Services (DTS). On October 1, 2001, the DTS

---

education and consumers. See "Convergence CCI to have Separate Bureaus on Content, Carriage" The Economic Times (20 July 2001) available at http://www.economictimes.com/200701/20tech28.htm

started the process of corporatizing to become an independent company. First, however, the government had to meet some of the demands of the employees who were against the corporatization of the DTS. To compensate the labour unions it promised to provide a free phone installation, free local calls to a certain number, and no rental charge for a telephone until a certain time.[30] The new company, which is wholly owned by the government, is called Bharat Sanchar Nigam Limited (BSNL). It has 400 000 workers.

The group also asked TRAI to make recommendations for allowing more competition in basic, cellular and long-distance services. While TRAI was having consultations before making recommendations, on August 9, 2000 the government relaxed strict conditions of sale of equity stakes between foreign and domestic investors in the industry. The cap of maximum 49 per cent of foreign equity is retained; however, this relaxation was seen as a major development for the expansion of the industry. Once this restriction was relaxed, there was a spree of mergers and takeovers in the industry.

The new policy allowed the majority held government companies MTNL and BSNL to compete as a third cellular operator with two private operators. From February 7, 2001, MTNL started offering cellular service in Delhi and Bombay. Even after reducing its tariffs substantially, which the private operators immediately matched, MTNL has not been very successful in attracting many subscribers and as of July 2001, it was successful in signing only 23 135 subscribers.[31] BSNL is yet to offer cellular services using GSM technology. It is carrying trials and expects to start services in 600 cities by early next year.[32]

---

[30] "Indian Telecoms Workers Strike over Uncertain Future" *Total Telecom* (6 September 2000) available at http://www.totaltele.com

[31] For more details of MTNL's failure to successfully offer cellular services, see, Sudesh Prasad, "Dolphin Living Dangerously" (6 September 2001) available at http://www.voicedata.com/content/news/101090603.asp

Recognizing the growth of Internet Telephony the government has realized that a monopoly over international long-distance services could not be sustained for long. The government decided to advance its commitments made to the WTO by two years and decided to end VSNL's monopoly on April 2002 and thereafter allow competition. It has also decided to allow Internet telephony after April 2002. (it is banned until then).[33] To compensate VSNL for ending its monopoly sooner than scheduled, the government, as part of the compensation package, has granted a nation- wide license for Internet and a free license for long-distance services. Although, VSNL wants the government to permit it to compete in cellular and basic services, the government has rejected these requests on the grounds that state players cannot compete in the same business with other state players. VSNL had challenged this order in the courts, causing temporary roadblock in the bidding process of the fourth cellular license.

Before VSNL loses its monopoly, the government has decided to reduce its stake from 52 per cent to 26 per cent.[34] Bids for selling its stake were called; however, foreseeing intense competition in infrastructure and services, investors showed little

---

[32] "India's BSNL Launches Three-year Plan for Cellular Rollout" *Total Telecom* (17 January 2001) available at http://wwwtotaltele.com

[33] "India to Allow Internet Telephony in April 2002" *Total Telecom* (12 April 2001) available at http://www.totaltele.com

[34] "Government Invites Offers for VSNL Equity" *Total Telecom* (20 February 2001) available at http://wwwtotaltele.com

enthusiasm. Out of six serious bidders, three have already withdrawn their applications.[35]

**Competition in Services**

After having a consultation process for allowing competition in basic, cellular and long-distance service TRAI had recommended unlimited competition in basic and domestic long-distance service. However, taking into consideration spectrum limitations, it recommended a fourth cellular player be introduced in the 21 circles.

On August 15, 2000, the government accepted TRAI's recommendations for opening unlimited competition in domestic long-distance services.[36] So far, apart from the government's BSNL and VSNL companies, two private companies who received a green signal from the government are Bharati Group and Reliance Industries limited. These companies are dominant players in the basic, cellular and Internet and are extensively deploying fiber optic infrastructure in various states.

---

[35] The six applicants were Videocon, the Reliance Industries, Tata Group, Birla Group, Bharati-Singtel Group and BPL Communications. The Birla group, Bharati and Singtel Group and Videocon have withdrawn their applications.

[36] The license is non- exclusive and is for 20 years period, extendable for 10 years at a time, for inter-circle long distance operations within India. Foreign equity is limited to 49 per cent. The applicant should submit its role out plan in four phases including the coverage of uneconomic and remote areas. The applicant has to pay one time entry fee of Rs. 100 Crores and provide four bank guarantees of Rs. 100 crores each to be released on completion of each phase's rollout obligations. The applicant apart from entry fee described above should pay license fee in the form of revenue share at 10 per cent and contributions towards universal service obligations to a total cap of 15 per cent. Also the applicant has to pay fee for using spectrum as prescribed. The applicant is allowed to change the name of the company or licensee. The operator is allowed to enter suitable arrangements with other service providers to negotiate interconnection agreements. The operator will comply with regard to interconnection by TRAI's regulation under the TRAI Act. It is mandatory for fixed service operators, cellular mobile service operators, and cable service operators, to provide interconnection to the long-distance operator so that subscribers could have a free choice to make domestic and international calls. For more details see *Guidelines for Issue of License for National Long Distance Service*. Government of India, Ministry of Communications, Department of Telecom available at http://www.dotindia.com

In March 2001, the government called tenders to provide licenses for fourth cellular operator.[37] The bidders showed great enthusiasm to bid for the fourth license because the cellular market is rapidly growing after the government shifted operators to the new licensing regime. It is noted that India had almost five million subscribers in September of 2001, almost double at the same time last year.[38] However, with the controversy of the basic service licenses, which will be discussed below, the valuations of the cellular licenses has tumbled to almost half its value.[39] Nevertheless, the government had received 57 initial bids for 17 circles. On August 31, 2001, it completed its bidding process and announced that Bharati Group won eight licenses,[40] Hutchison Telecom won three licenses[41], Birla AT&T Communications won one license[42], Escorts

---

[37] The guidelines for issue of license for cellular mobile telephone service states that the bidder should be an Indian company with maximum 49 per cent of foreign equity. Bidders are allowed to bid separately for each service area (approximately a state) and can apply for any number of service areas. The license will be issued for 20 years extendable for 10 years each time. In its roll out obligations of its service area, the operator has to provide 10 per cent in the district head quarters in the first year and 50 per cent within in 3 years. In cities 90 per cent in the first year. The successful bidder will be required to pay one time entry fee based on the final bid before signing the license agreement. In addition it shall also pay maximum 18 per cent as a revenue share generated from service. This fee includes contribution for universal service obligations, research and development and for providing minimum spectrum. Each licensee is allowed to carry its subscriber's traffic in it own service area. Interconnectivity among service providers within in same area is permitted by mutual agreements. The cellular operators may enter mutual agreements with other service providers for terminating traffic in other service areas. Interconnection agreements are also subject to regulations issued by TRAI under the TRAI Act. It is mandatory for cellular operators to provide interconnection to long-distance carriers so that subscribers may have a free choice to make domestic long distance and international calls. However, for international long-distance call, the cellular operator has access national long distance company only (BSNL). Resale/assignability/ transferability is permitted subject to approval. The *guidelines for Fourth Cellular Licensing* can be found at www.http://dotindia.com

[38] "India's Mobile Users Hit 4.5 Million" (19 September 2001) available at http://www.cnnasia.com

[39] For example the Chennai service area valuation is believed to have almost halved to a modest $75 million- $80 million from $ 150 million. See "Hutchison to Bid for More Cellular Licenses" *Total Telecom* (22 June 2001) available at http://www.totaltele.com

[40] Maharashtra, Gujarat, Bombay, Tamil Nadu, Haryana, Kerala, Western Uttar Pradesh and Madhya Pradesh.

[41] Madras and Karnataka

[42] New Delhi

Telecommunications won four licenses[43] and the Reliance Group won one license.[44] The government is set to earn $ 339.5 million from the entry fees.[45]

**Controversy regarding Basic Service Licenses**

With the failure of the first round of reforms to spread basic telephony teledensity increased from 0.8 in 1994 to 3 per cent until early 2001. The government's target is to have 7 per cent by 2005 to comply with its 99 policy commitments. Therefore, on 23 April 1999, DOT asked TRAI to recommend license conditions for basic service providers (BSPs). Consequently, TRAI invited various stakeholders to discuss various issues relating to licenses of BSPs. It also suggested that the wireless local loop (WLL) technology, which allows limited mobility with handsets, should be permitted for faster rollout by BSPs.[46] However, this suggestion was objected by the Cellular Operators Association (COA), who suggested that the "poor man's mobile phone" based on WLL technology for a short distance for achieving universal access goals was nothing but a substitutable cellular service. The COA stated that if WLL were allowed, it would amount to a "back door entry" for the BSPs to compete with cellular operators with unfair terms. They said this decision would disrupt the level playing field as cellular operators compared to BSPs pay far higher fees for entry, revenue share and spectrum.

---

[43] Punjab, Himachal Pradesh, Uttar Pradesh (east) and Rajasthan

[44] Calcutta

[45] "Bharti Sweeps Indian Mobile License Auction" Total Telecom (31 July 2001) available at http://www.totaltele.com

[46] The Wireless Local Loop technology is India's indigenous technology. It was highly appreciated by UNDP. This technology replaces the wires and copper in local loop with a wireless system. It requires a compact base station, mounted on a rooftop or poles in the streets, which is used for transmission on a wireless medium to homes and offices. The technology is claimed to bring down the cost of per line telephone connection from Rs. 40 000 to 10 000 and facilities both voice and data. It is suited for both dense urban and sparse rural deployment scenarios. India and other nations such as Fiji, Tunisia, Nigeria and Madagascar have already started to deploy. See "India's WLL Tech Praised By UNDP" *The Economic Times* (1 August 2001) available at http://www.economictimes.com/010801/01tech04.htm

However, on 8<sup>th</sup> January 2001, TRAI recommended for BSPs that it was not treating the provision of limited mobility with WLL as a service outside the ambit of their service provision. It said to do otherwise would be to prevent consumers from benefiting from the fruits of the technological progress. It noted that the quality of service provided by cellular operators was superior to what will be provided by BSP's by using WLL and, hence, it will not effect the cellular operators' business, it also stated that it is a different service. It said it views WLL with limited mobility similar to a supplementary or value-added service for basic service. In that sense, this service would be similar to the supplementary services and roaming services that are presently allowed for cellular mobiles. The Authority further said that there is no reason to reconsider the issue of an entry fee for BSPs, particularly because the purpose of an entry fee was mainly to deter non-serious entry of service providers. Likewise, the license fee and revenue share percentages need not be altered for BSPs. Though their revenue streams will now be higher, the amount of revenue share license fee will also be higher as a consequence. The Authority does not favour imposing a greater license fee burden on the service providers, as it will pass these high fees on to the consumer. It also said the charge made from WLL handsets should be same as the local call set at Rs. 1.20 per 3 min.

The COA reacted to the recommendations of TRAI, saying they were "extremely sketchy" and that they were issued with a "pre-determined mind" and they said they would hold back their investments for the fourth cellular license because of "commercial and regulatory instability."[47] On January 22, 2001, they filed a petition in TDSAT seeking that DOT not to consider the recommendations of TRAI. However, on January 24, 2001, DOT went ahead and accepted TRAI 's recommendations and issued guidelines for issuing basic service licenses; it also stipulated that the allocation of spectrum is free and on "first came first served" basis for WLL operations. [48]

The announcement of free spectrum on first come, first served basis further led to a heated clash between the basic and cellular associations. The COA requested, that the Prime Minister intervene to safeguard their interest as the Minister of Communications had provided them with a deaf ear and was totally in favour of TRAI's recommendations. Moreover, there was no concrete decision from TDSAT, which only stated that any license issued would abide by the result of the petition filed by the COA.

With no clarity on commercial and regulatory issues, both cellular and basic service operators rushed to apply for fixed service licenses as the spectrum was free and was allotted on a first come, first served basis. By February 9, 2001, DOT received 132 applications and it quickly evaluated them. On March 26, 2001, it issued letters of intent (LOI) to 40 applicants (15 to M/S Tata Teleservices, 18 of M/S Reliance Communications and 7 of M/S HFCL Infotel).

---

[47] "India's COAI Declares War after WLL-based Mobility Ruling" Total Telecom (10 January 2001).

[48] On 25 January, 2001, the government issued guidelines to issue basic service licenses, which stated that, will be no restriction on number of operators to be allowed to offer basic telephone service in each state. Company's can apply for any number of circles making separate applications for each circle. The license issued will be on non-exclusive basis for a period of 20 years and extendable for 10 years at one time. Foreign equity to be restricted to 49 per cent. The applicant has to spread telephony in its circle in four phases as prescribed. The company has to pay one time prescribed entry fee prescribed for each circle. There will be no separate entry fee payable for allocation and usage of spectrum. In addition to entry fee described, there will be license fee of maximum 12 per cent from the gross revenue depending on the telecom circle the operator operates. And additional revenue of 2 per cent from gross revenue earned from WLL subscribers shall be levied as spectrum charge for allocation of 5 plus 5 Mhz. Allocation of spectrum to basic service operators will be allowed on first come first basis. Basic service operators will be allowed to provide mobility to its subscribers within local area restricted to 10 km. Cellular operators is also provided to offer basic services in their circle using their GSM network. Calls for WLL subscribers may be charged at Rs. 1.20 per unit call (3 min). Rental for WLL subscribers shall be fixed by TRAI. Interconnection with other networks shall be based on mutual agreements between service providers or directions from TRAI under TRAI Act. However, for international long distance calls, the basic service operator shall access international long-distance operator through national long-distance operator only. Fore more details see Guidelines for issue of license for Basic Service available at http://www.dotindia.com

However, the controversy which was brewing for a year and half finally got attention from the Prime Minister's Office when the opposition political parties alleged that this was another telescam of Rs.13 crores favouring basic service operators. Consequently, the government immediately asked DOT to stop issuing LOI's and seized the matter, deciding to make the final decision by April 30th. It then referred the matter to the GTC to resolve the problem. The GTC's focus was to see if NTP-99 permitted "limited mobility" service to be offered by BSP's, and, if NTP-99 did not allow it, to determine how the policy could be altered to include this service so that it can help in achieving teledensity targets at cheaper and affordable rates. If the policy did allow limited mobility for BSP's, the GTC wanted to explore how it could be introduced to be consistent with the principle of a level playing field and different categories of operators with the objective of assured services at the cheapest possible rates.

At the outset, the GTC made it clear that all of these issues fall squarely within the jurisdiction of TRAI and TDSAT. However, since the government had asked their assessment on specific issues it would give its recommendations.

It came to the conclusion that NTP- 99 allowed the use of WLL by the BSPs. The GTC noted CAO's contention that the two services will be substitutable, since BSP's will offer services at local call rates, and that the CMO's will face unequal and unfair competition which will disturb level playing field. On this point the group did not accept the recommendation of TRAI and came to a conclusion that to ensure fair competition the present revenue sharing arrangement between the BSP's and the long-distance carriers was 60:40 was not fair if BSPs want to provide WLL. To create a level playing field, BSP's should share in the ratio of 5:95 in the same way that the CMO's presently share with their long-distance providers.

However, the group did not make any decision on a separate entry fee for utilizing WLL and referred the matter back to TRAI to decide. It also noted that the description of "first come, first served" used in the guidelines was not accurate with regard to spectrum allocation.[49] It directed DOT to follow TRAI's recommendation on March 23, 2001 which grants spectrum free,[50] but provides detailed conditions under which spectrum would be allotted.

The government endorsed the recommendations made by GTC. TRAI later fixed rates for the phone rentals at Rs. 450.COA's reaction was that the "poor man's cellular phone" service permitted for BSP's would be unaffordable because of the cost of handsets and the rentals of the phone. COA's did not withdraw its petition from TADSAT and matters are still pending in the Tribunal

**Conclusion**

All these developments show that India has come a long way, from a very closed economy to a more decentralized model. The government's reluctance in the first round of reforms to break the dominance of DOT and to overhaul the legal and regulatory regime led to endless litigation, which delayed the liberalization process for almost a decade.

---

[49] Report of the Group on Telecom & IT Convergence on Limited Mobility is available at http://www.dotindia.com

Explosive growth of the Internet and wireless technologies and the threat by investors to decamp and withdraw their investment led the government to make decisions to untangle the problems. The *1999 Telecom Policy*, the *Internet Policy*, and the recent legal and regulatory initiatives to overhaul the legal and regulatory regime are all steps in a positive direction. Competition in basic, cellular and domestic long-distance services will also help in the faster rollout of telephony. The government will have to make some shrewder decisions in privatizing its incumbent companies to create a level playing field. Separation of DOT policy making role and creating BSNL for providing services is a positive move. However, we have seen that DOT continues to view BSNL as a national incumbent company. The new licensing regime of a one- time entry fee and rest by revenue sharing is more practical and reasonable than the previous high license fee regime. Nevertheless, in the licenses to the new entrants the DOT has again taken a protectionist stand by asking them to pass all international traffic only through the national long-distance company BSNL. This is contrary to the government's stand in which it stated that BSNL is an independent company and will be treated the same as other companies.

The decision to allow WLL by the TRAI to achieve universal access goals was a positive decision, especially when the technology is indigenously developed, because it can tremendously cut the cost and help in spreading telephony at a faster pace. However, the WLL controversy also highlights that it is important for the policy makers to make mature decisions and to create a level playing field to avoid adverse effects to the liberalization policy.

---

[50] Government of India, Ministry of Communications, Department of Telecommunications, "Procedure for Allocation of Spectrum on First Come First Served Basis" (23 March 2001) available at

Since India has allowed competition in various services, is it not time for the policy makers to consider issuing a single license for offering various services with minimum requirements while framing the new legislation.

The overall liberalization trend shows that, subject to these lingering transition problems, the government's mindset has significantly moved from a closed regime to a pro-competitive regime. In this round of reforms the government is much more ready to partner with various public and private interest groups to achieve public policy goals.

http://www.dotindia.com